\documentstyle[prl,aps,multicol,graphicx]{revtex}

\newcommand{\be}{\begin{equation}}
\newcommand{\ee}{\end{equation}}

\newcommand{\ba}{\begin{eqnarray}}
\newcommand{\ea}{\end{eqnarray}}

\begin{document}

\title{An Improved Upper Bound for the Ground State Energy\\
of Fermion Lattice Models}
\author{Matteo Beccaria}
\address{
Dipartimento di Fisica, Universit\`a di Lecce and INFN, Via Arnesano,
73100 Lecce, Italy }
\date{\today}
\maketitle
\begin{abstract}
We present an improved upper bound for the ground state energy of 
lattice fermion models with sign problem. The bound 
can be computed by numerical simulation of a recently 
proposed family of deformed Hamiltonians with no sign problem. 
For one dimensional models, 
we expect the bound to be particularly effective and practical extrapolation procedures
are discussed.
In particular, in a model of spinless interacting fermions and in the Hubbard model at
various filling and Coulomb repulsion we show how such techniques can
estimate ground state energies and correlation function with great accuracy.
\end{abstract}

\begin{multicols}{2}
\narrowtext


In the study of strongly interacting quantum systems, the 
determination of ground state properties is a basic step in the analysis of
many physically relevant problems.
To accomplish this task, several powerful numerical methods are available 
and big efforts have been spent in 
their development during the last years~\cite{Linden}.
However, as is well known, in fermion models, numerical
simulations 
face serious obstructions that dramatically  reduce their 
performance~\cite{Loh}.
In this Brief Report, we propose an improved upper bound for the 
ground state
energy and an effective strategy to
circumvent this difficulty in a certain class of one
dimensional models.

For moderate lattice size, Exact Diagonalization 
methods are possible~\cite{Dagotto} with no additional problems in the
fermionic case. Typically, these are 
calculations based on the Lanczos algorithm and the ground
state of systems with up to several thousands states can be determined.  
For larger state space, other methods must be considered. 
In particular, the most flexible tool is Monte Carlo 
simulation that evaluates quantum expectation values
by a careful stochastic sampling of the configuration space.

Often, when the model under study is fermionic,  the measure to
be sampled is non-positive.
This fact can be related to the intrinsic anticommuting nature of the Fermi 
creation annihilation operators, but is more general and appears also in
bosonic contexts, like quantum spin models with non trivial
exchange terms~\cite{Henelius}.
This situation, usually described as the ``sign-problem'', is
quite serious and standard algorithms simply fail.

Several proposals are available to overcome or at least reduce the sign
problem in specific problems, in particular
realistic models in more than one dimension. One of the most powerful
techniques is the Constrained Path Monte Carlo Method~\cite{Gubernatis}
where the sign problem
is eliminated by introducing a physically motivated guiding wave function 
that approximates the nodal structure of the exact ground state.
A numerical analysis is then possible and expectation values can be
computed. The approximation is partially controlled and, for instance,
the estimates of the ground state energy are known to provide rigorous upper bounds.
Other alternatives are the Fixed-Node approximation, the Projection Quantum
Monte Carlo, path integral representations or the Auxiliary Field Monte 
Carlo~\cite{methods}.
Of course, trial wave functions are useful also in cases with no sign
problem, but their purpose to accelerate Monte Carlo simulations
and the final results are in principle 
independent on the choice of the guiding function.


Our first goal will be that of providing an 
improved general upper bound for the energy 
of fermion lattice models with sign problem; the bound is computable by numerical
simulations.

As a second aim, we ask whether one dimensional problems allow for any 
special simplification and practical treatment of the
sign problem in the spirit of~\cite{alhassid}. 
In particular we look for methods that do not rely on 
any biased approximation.
A hint to a positive answer comes from the remark that in one dimension, 
the sign problem appears to be somewhat artificial.
For instance, we can consider a model where the standard fermion hopping 
term is the unique source of alternating signs in the off diagonal matrix 
elements of the Hamiltonian.
In this case, the sign problem disappears as soon as open boundary
conditions are used~\cite{one}.
Its manifestations are therefore expected to be completely 
negligible in observable quantities that admit a thermodynamical limit,
independent on the boundary conditions. Possible examples are
the ground state energy per site and finite correlation lengths.
For this reason, we are led to believe that in such cases, 
the finite size effect of the sign problem should be 
under control without much effort.

To begin our analysis, 
let us consider a $d$ dimensional hypercubic lattice with $V=L_1L_2\cdots
L_d$ sites and periodic boundary conditions.
Let us assign to each site $\bf i$ a pair of spinless fermion creation
annihilation operators $c_{\bf i}$ and $c^\dagger_{\bf i}$
with canonical algebra
\be
\{c_{\bf i}, c^\dagger_{\bf j}\} = \delta_{{\bf i},{\bf j}},\qquad
\{c_{\bf i}, c_{\bf j}\} = 0,\qquad
\{c^\dagger_{\bf i}, c^\dagger_{\bf j}\} = 0 .
\ee
Let us denote a configuration by $\bf n$, the set of occupation numbers at
all the lattice sites. 
We work in the occupation number basis with states $\{|{\bf n}\rangle\}$
such that $c^\dagger_{\bf i} c_{\bf i} |{\bf n}\rangle = n_{\bf i}
|{\bf n}\rangle$ and assume a given ordering of the sites to fix signs.
We want to study Hamiltonians of the form 
\be
\label{ham}
H = H_{kin}+H_{int},
\ee
where 
\be
H_{kin} = -t \sum_{\langle {\bf i}, {\bf j}\rangle}
 (c_{\bf i}^\dagger c_{\bf j} +c_{\bf j}^\dagger c_{\bf i}),
\ee
($\langle {\bf i}, {\bf j}\rangle$ denotes 
a pair of connected sites, for instance 
nearest neighbours). The 
interaction term is taken to be diagonal in the $\{|{\bf n}\rangle\}$ basis
\be
H_{int} = V({\bf n}), \qquad n_{\bf i} = c_{\bf i}^\dagger c_{\bf i}, 
\ee
with $V$ being a real function.
The fermion number $N = \sum_{\bf i} n_{\bf i}$ is conserved and the analysis 
of the spectrum can be carried on in each sector with definite $N$.

In the one dimensional case, $V=L_1\equiv L$, we can assume the 
following canonical site ordering 
\be
|{\bf n}\rangle = (c_1^\dagger)^{n_1}\cdots(c_L^\dagger)^{n_L} |{\bf
0}\rangle ,
\ee
where $|{\bf 0}\rangle$ is the empty state. The paired sites are
$\langle i, i+1\rangle$ modulo periodic boundary conditions.
The off diagonal matrix elements of $H$ are then negative and Monte Carlo 
simulations do not have sign problems, except when the
fermion number $N$ is even; in this case, there exist matrix elements with
the wrong sign, precisely those connecting pairs of states that differ
by a hopping through the boundary.

As shown in~\cite{Ceperley}, it is possible to introduce a new Hamiltonian,
$H^{\rm eff}$, with no sign problem. Its ground state energy $E_0^{\rm eff}$ can be
computed by numerical methods and provides a rigorous upper bound for the 
ground state energy $E_0$ of $H$. In~\cite{Sorella2}, this result has been
extended by introducing a family of 
Hamiltonians $\{H(\gamma)\}$ depending on a real free parameter $\gamma$. 
Let us recall the definition of
$H(\gamma)=\{H_{ij}(\gamma)\}$ in terms of $H=\{H_{ij}\}$.
We assume $H=DH^{(0)}D^{-1}$ with $H^{(0)}$ hermitean (for instance of the
form Eq.~(\ref{ham})) and $D$ diagonal with $D_{ii}\neq 0$. $D$ is a
trial wave function useful to accelerate convergence to the ground
state. The new Hamiltonian $H(\gamma)$ is given by  
\be
H_{ij}(\gamma) = \left\{
\begin{array}{lll}
H_{ij} & H_{ij}\le 0 & i\neq j \\
-\gamma H_{ij} & H_{ij}>0 & i\neq j \\
H_{ii}+(1+\gamma)\sum_{H_{ik}> 0, k\neq i} H_{ik} &  & i=j
\end{array}
\right. .
\ee
If
we call $E_0(\gamma)$ the ground state energy of $H(\gamma)$, then it is
possible to show that: (i) $E_0(-1)=E_0$, (ii) $E_0(\gamma)\ge E_0$ for any
$\gamma\ge -1$, (iii) $H(\gamma)$ has no sign problem if $\gamma\ge 0$. 
In the Monte Carlo approach adopted in~\cite{Sorella2}, it is also emphasized that 
a strictly positive
$\gamma$ must be used and that an extrapolation to $\gamma=0$ is required
for best accuracy.

We now show that the function $E_0(\gamma)$ is concave in
$\gamma$. Namely, for $0\le \alpha\le 1$ and any two real $\gamma_1$,
$\gamma_2$,  we have 
\be
E_0(\alpha\gamma_1+(1-\alpha)\gamma_2) \ge \alpha E_0(\gamma_1)+ 
(1-\alpha) E_0(\gamma_2) .
\ee 
A general simple proof can be given by means of the variational
characterization of $E_0$. 
The dependence of $H(\gamma)$ on the parameter 
$\gamma$ is linear and we can write
\be
H(\gamma) = D H^{(0)}(\gamma) D^{-1} = D(H_1+\gamma H_2)D^{-1} ,
\ee
with hermitean, $\gamma$ independent $H_1$ and $H_2$. Then, if $S$ is the set of normalized stated,
we have
\ba
\lefteqn{E_0(\alpha\gamma_1+(1-\alpha)\gamma_2) = }\nonumber\\
&& \min_{s\in S} \langle s | \alpha H^{(0)}(\gamma_1)+(1-\alpha) H^{(0)}(\gamma_2)
|s\rangle \ge  \\
&& \ge 
\alpha\min_{s\in S}
\langle s|H^{(0)}(\gamma_1)| s\rangle +
(1-\alpha)\min_{s\in S}
\langle s|H^{(0)}(\gamma_2) | s\rangle = \nonumber\\
&& \alpha E_0(\gamma_1)+
(1-\alpha)E_0(\gamma_2).
 \nonumber
\ea 
Concavity means that the incremental ratio is a decreasing function.
Computing the incremental ratio on the intervals $[-1,0]$ and $[\gamma_1, 
\gamma_2]$ with arbitrary $0 \le \gamma_1 < \gamma_2$ we find
\be
E_0 \le  E_0(0)-\frac{E_0(\gamma_2)-E_0(\gamma_1)}{\gamma_2-\gamma_1}.
\ee
If we furthermore assume $E_0(\gamma)$ to be differentiable at $\gamma=0$
and take the limit $\gamma_2, \gamma_1\to 0$ we can replace the 
the upper bound  $E_0\le E_0(0)$ by the 
the improved one
\be
E_0 \le E_0(0)-E_0'(0^+) ,
\ee
that expresses the geometrical fact that a concave curve lies below its
tangent at any point. 
We remark that $E_0'(\gamma)$ can have finite jumps. 
If they can be excluded a priori, 
than the above proof follows also by using second order perturbation theory.

The concavity constraint suggests that the function $E_0(\gamma)$
could be particularly well behaved. For this reason we
also try to obtain an extrapolation of its value at $\gamma=-1$ by assuming
it smooth enough. To be fair, this possibility must be considered just a
practical recipe. 
Nonetheless, we shall give numerical arguments to support its robustness.

We begin with a simple model of interacting spinless fermions. The
Hamiltonian is 
\ba
\label{spinless}
\lefteqn{H = \sum_i\left[-t(c^\dagger_i c_{i+1} + c^\dagger_{i+1}
c_i)\right.+} && \\ 
&&\qquad  +\left.  V \left(n_i-\frac 1 2\right)\left(n_{i+1}-\frac 1
2\right)\right].
\nonumber
\ea
For lattice size $L=8$, $12$, $16$ and $20$, at half filling 
and for several interaction couplings $V$, we determine by Lanczos
diagonalization the exact ground state energy $E_0$, the bound $E_0(0)$,
the derivative $\eta = E_0'(0)$ and the improved bound
$E_0^{(imp)} = E_0(0)-\eta$. Moreover, we also attempt an extrapolation to $\gamma=-1$
starting from the values of $E_0(\gamma)$ in the range $\gamma\in(0,1)$. 
In principle, any such extrapolation can be very dangerous. On the other
hand, the extrapolated $E_0^{(m)}(-1)$ obtained by a polynomial
fit of degree $m$ displays a flat behaviour with small oscillations unless
$m$ is too large when the extrapolation process breaks down. 
The typical degree of the best polynomial is $8$.
To accelerate the convergence and to determine the best
estimate, we also used Aitkin's algorithm~\cite{acc} that improves the 
converging sequence $\{z_n\}$ replacing it with $\{z'_n\}$ defined by  
\be
z_n' = z_n-\frac{(\Delta z_n)^2}{\Delta^2 z_n},\quad \Delta z_n =
z_{n+1}-z_{n} ,
\ee
(where $\Delta^2 z_n = \Delta(\Delta z_n)$).
We call
$E_0^{(ext)}$ the resulting prediction. 
In Fig.~1,
we show the results for $E_0$, $E_0(0)$, $E_0^{(imp)}$ and $E_0^{(ext)}$ in
terms of their relative percentual accuracy defined as $100\cdot 
|\delta E/E_0|$
where $\delta E$ is the error in the ground state energy estimate.
The improved bound is significantly better than $E_0(0)$. Both $E_0(0)$ and $E_0^{(imp)}$ converge to 
the exact value as the system size $L\to \infty$ as expected from our initial discussion on the 
asymptotic irrelevance of the boundary effects. The extrapolated bound $E_0^{(ext)}$ is quite 
precise and around the permille level.

A similar analysis can be done for the measurement of
observables. Unfortunately, for these, we cannot derive a simple bound like
the one for the energy and in fact random hermitean operators produce 
easily wild behaviour for $\gamma\in(-1,0)$. On the other hand, for 
operators associated to physically meaningful quantities independent on the 
boundary conditions we can expect a mild dependence on $\gamma$. 
As a typical non local observable, we consider the
integrated staggered correlation function defined by 
\be
S = \frac 1 L \sum_{i=1}^L\sum_{k=1}^{L/2} (-1)^k ( \langle n_i n_{i+k}\rangle -\langle n\rangle^2).
\ee
In Fig.~2, we show how the exact values are very accurately reproduced
by the extrapolated values. The relative accuracy is always well below the
percent level.
In Fig.~3, we show the behaviour of the functions $E_0(\gamma)$ and 
$S(\gamma)$ at $V=1$ for the system with $L=16$ together with a 8-th degree
polynomial fit to emphasize their smoothness.

For large systems, the upper bound must be computed by numerically extrapolating 
$E_0(0)$ and $E_0'(0)$. This can be done by a straightforward Monte Carlo simulation
at variable $\gamma$. To explore the feasibility of this proposal, 
we perform such a study for free fermions ($V=0$) on a lattice
with $L=40$ sites at half filling. Here, the exact ground state energy is known, 
$E_0/2t = -\cot(\pi/L)\simeq -12.706$ and can be used as a check. We use the
Green Function Monte Carlo
 with Stochastic Reconfiguration~\cite{SorellaGFMC,sun}. 
We run simulations
with variable number of walkers in order to extrapolate  the infinite population size limit.
In Fig.~4 we show the estimated value of $E_0(\gamma)$ at several positive $\gamma$ values
together with a simple parabolic fit. The estimated value of the bound is $E_0(0)-E_0'(0) = -12.61(1)$,
about $1\%$ off the exact value.

Similar results are obtained by studying the one dimensional Hubbard
model. Denoting by $\uparrow$, $\downarrow$ the two spin degrees of
freedom, the Hamiltonian reads
\be
\label{Hubbard}
H = -t \sum_{i,\sigma=\uparrow, \downarrow}(c^\dagger_{i, \sigma}
c_{i+1, \sigma} + c^\dagger_{i+1, \sigma} c_{i, \sigma})
+ U \sum_i n_{i, \uparrow} n_{i, \downarrow} .
\ee
Again, we determine for several lattice size, filling fraction and
coupling $U$ the four quantities $E_0$, $E_0(0)$, $E_0^{imp}$ and
$E_0^{ext}$. The results are shown in Fig.~5 where we show the relative percentual
accuracy of $E_0^{(ext)}$. The quality of the results is not good as in the spinless model, but
the errors are again small, of a few percents.  In fact, a scaling analysis shows convergence to 
the exact values in the large volume limit.

To summarize,  
the family of Hamiltonian with no sign problem proposed
in~\cite{Sorella2} makes possible the derivation of a size consistent bound
for the ground state energy that improves the one at $\gamma=0$.
Moreover, much information can be reconstructed for 
the original Hamiltonian.
The accuracy of our analysis on small systems is
certainly beyond a practical implementation, but suggests that also for more
complicated systems, not allowing a direct analysis, 
the extrapolated values can provide useful
numerical hints. 
Preliminary results on the two dimensional Hubbard model are
encouraging and will be presented elsewhere~\cite{prep}.

Partial support of INFN, Iniziativa Specifica RM42, is acknowledged.


\begin{figure}
\centerline{\includegraphics*[width=6.5cm,angle=-90]{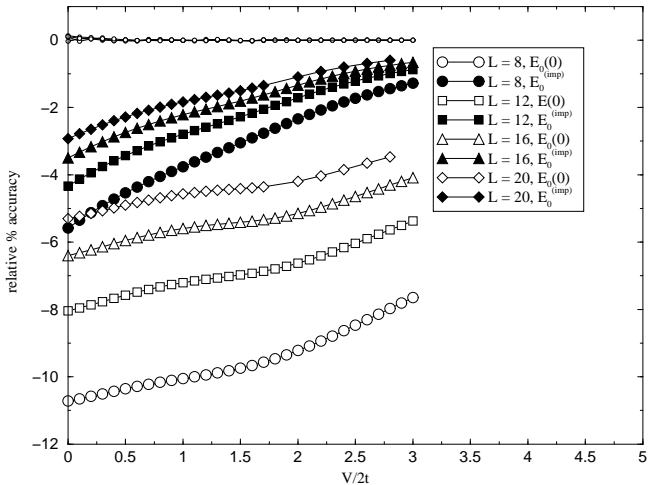}}
\caption{
Spinless interacting fermions. 
Relative percentual accuracy of the ground state energy bounds and extrapolated estimate.
The circles with very small error are the best extrapolated 
bound $E_0^{(ext)}$ at the various lattice sizes considered $L=8, 12, 16, 20$
(we use the same symbol since the aim is to emphasize the small maximum
deviation). 
}
\end{figure}\noindent

\begin{figure}
\centerline{\includegraphics*[width=6.5cm,angle=-90]{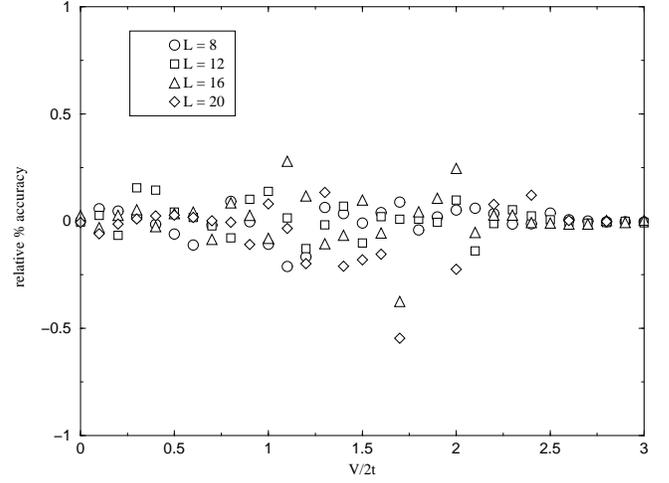}}
\caption{Spinless interacting fermions. 
Relative percentual accuracy of the extrapolated estimate of the integrated 
staggered correlation function $S$.
}
\end{figure}\noindent

\begin{figure}
\centerline{\includegraphics*[width=6.5cm,angle=-90]{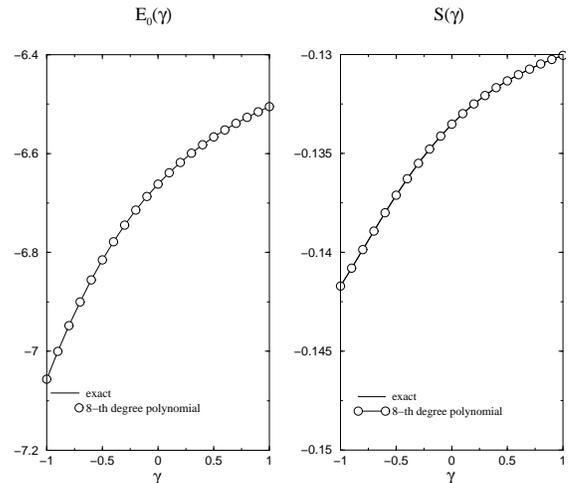}}
\caption{Spinless interacting fermions. 
Smoothness of the function $E_0(\gamma)$, $S(\gamma)$ and 
accuracy of a 8-th degree polynomial fit for the model at $V=1$ with 
$L=16$ sites.
}
\end{figure}\noindent

\begin{figure}
\centerline{\includegraphics*[width=6.5cm,angle=-90]{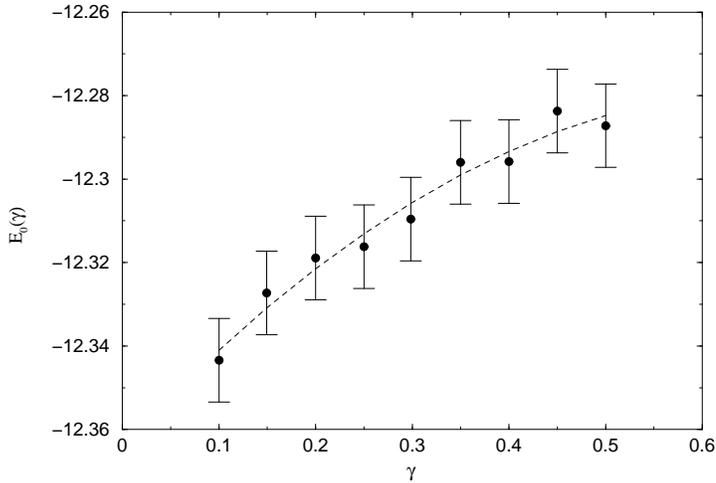}}
\caption{Spinless interacting fermions.
GFMC calculation of the effective energy
$E_0(\gamma)$ in the free case $V=0$, $t=0.5$,  on a lattice with $L=40$
sites at half filling. Data are obtained by the GFMC  
with Stochastic Reconfiguration 
in the Poisson limit (continuous time). The size of the 
walker ensemble has been varied from $300$ up to $2000$ and the extrapolation
to the infinite population size has been taken.
Statistics corresponds to $10^5$ reconfigurations separated by an evolution
time $\tau=0.5$. The dashed line is a parabolic fit.
}
\end{figure}\noindent

\begin{figure}
\centerline{\includegraphics*[width=6.5cm,angle=-90]{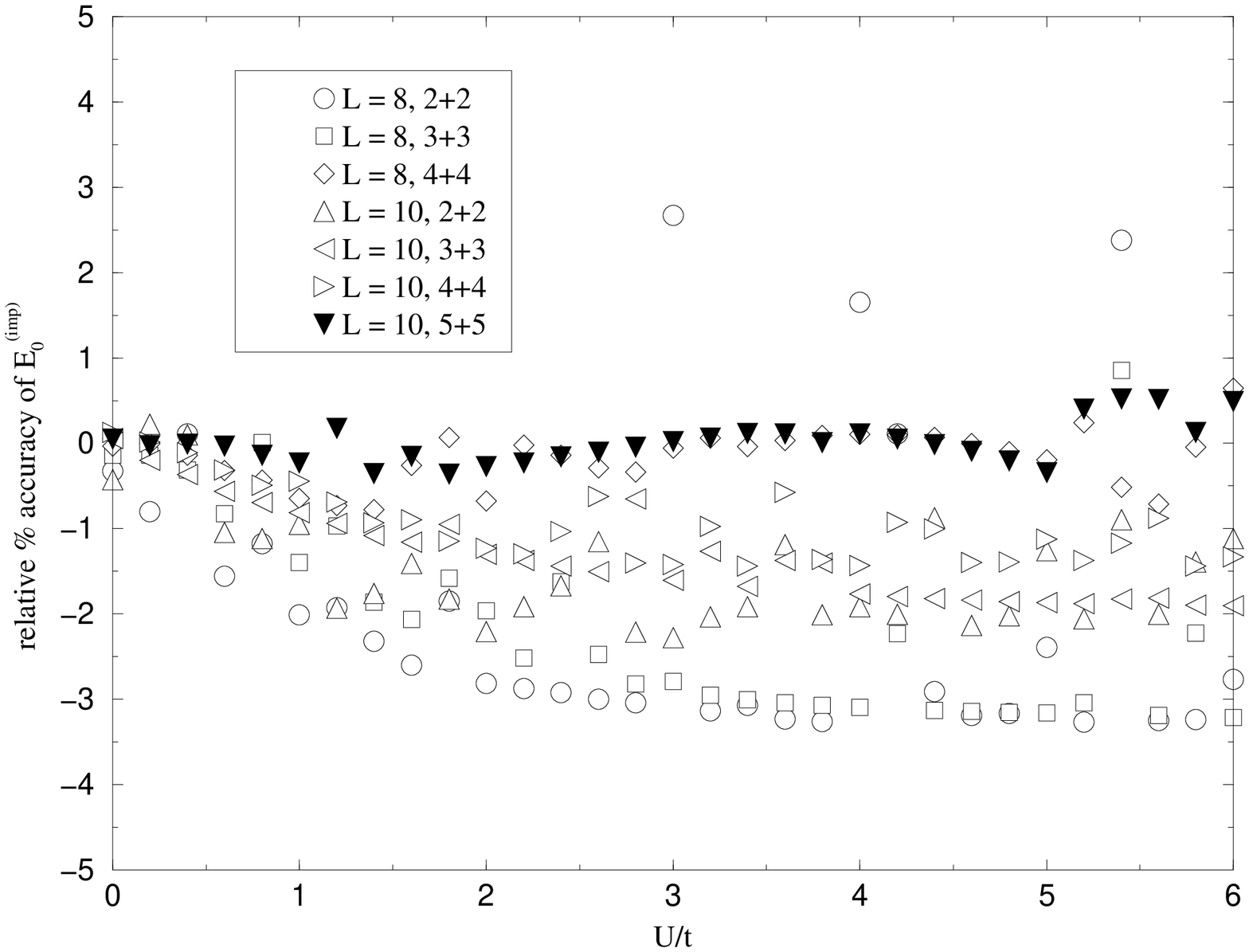}}
\caption{Hubbard model. Relative percentual accuracy of the extrapolated 
estimate of the ground state energy at different sizes, filling and Coulomb coupling.}
\end{figure}\noindent

\end{multicols}

\begin{references}



\bibitem{Linden} 
W. von der Linden, Phys.~Rep.~{\bf 220}, 53 (1992).  

\bibitem{Loh}
E.~Y.~Loh, J.~E.~Gubernatis, R.~T.~Scalettar, S.~R.~White, 
D.~J.~Scalapino and R.~L.~Sugar, Phys.~Rev.~B{\bf 41}, 9301, (1990).



\bibitem{Dagotto}
E.~Dagotto, Rev.~Mod.~Phys.~{\bf 66}, 763, (1994). 
For applications in one dimension, mainly with open boundary conditions, 
a powerful technique is the Density Matrix Renormalization Group, 
S.~R.~White, Phys.~Rev.~Lett.~{\bf 69},2863, (1992);
S.~R.~White, Phys.~Rev.~B{\bf 48}, 10345, (1993).

\bibitem{Henelius}
P.~Henelius and A.~W.~Sandvik, Phys.~Rev.~B{\bf 62}, 1102, (2000).

\bibitem{Gubernatis}
J.~Carlson, J.~E.~Gubernatis, G.~Ortiz, S.~Zhang, 
Phys.~Rev.~B{\bf 59}, 12798, (1999).

\bibitem{methods}
J.~B.~Anderson, J.~Chem.~Phys.~{\bf 65}, 4122, (1976);
G.~Sugiyama, S.~.E.~Koonin, Ann.~Phys.~(NY){\bf 168}, 1, (1986);
D.~M.~Ceperley, Rev.~Mod.~Phys.~{\bf 67}, 279, (1995);
R.~Blanckenbecler, D.~J.~Scalapino, R.~L.~Sugar, 
Phys.~Rev.~D{\bf 24}, 2278, (1981); Phys.~Rev.~D{\bf  24}, 4295, (1981).


\bibitem{alhassid}
Y.~Alhassid et al., Phys.~Rev.~Lett.~{\bf 72}, 613, (1994).



\bibitem{one}
J. Hirsch, D. J. Scalapino, R. L. Sugar and R. Blankenbecler,
Phys.~Rev.~Lett.~{\bf 47}, 1628 (1981);
J. Hirsch, D. J. Scalapino, R. L. Sugar and R. Blankenbecler,
Phys.~Rev.~B{\bf 26}, 5033, (1982).


\bibitem{Ceperley}
D.~F.~B.~ten Haaf, H.~J.~M.~van Bemmel, 
J.~M.~J.~van Leeuwen, W.~van Saarlos and D.~M.~Ceperley, Phys.~Rev.~B{\bf 51},
13039, (1995).


\bibitem{Sorella2}
S.~Sorella and L.~Capriotti, Phys.~Rev.~B{\bf 61}, 2599, (2000).

\bibitem{acc}
A. C. Aitken, Proc.~Roy.~Soc.~Edinburgh~{\bf 46}, 289 (1926);
C. Brezinski, Acc\'el\'eration de la convergence en Analyse Num\'erique, Springer-Verlag, Berlin, 1977.

\bibitem{SorellaGFMC}
S.~Sorella, Phys.~Rev.~Lett.~{\bf 80}, 4558, (1998);
N.~Trivedi and D.~M.~Ceperley, Phys.~Rev.~B{\bf 41}, 4552,
(1990). For a discussion of its continuous evolution time limit 
see M.~Beccaria, C.~Presilla, G.~F.~De~Angelis, G.~Jona-Lasinio,
Europhys.~Lett.~{\bf 48}, 243, (1999);
%



\bibitem{sun}
C.~J.~Hamer, M.~Samaras and R.~J.~Bursill, Phys.~Rev.~D{\bf 62}, 074506, (2000);
M.~Beccaria, Phys.~Rev.~D{\bf 62}, 034510, (2000).

\bibitem{prep}
M. Beccaria, A. Moro, in preparation.



%
%
%
%
%
%
%
%
%
%
%
%



\end{references}
\end{document}